\documentclass[twocolumn,showpacs,preprintnumbers,amsmath,amssymb,
superscriptaddress]{revtex4}

\usepackage{graphicx}
\usepackage{dcolumn}
\usepackage{bm}

\begin{document}

\preprint{}

\title{Superfluid Density and Field-Induced Magnetism in Ba(Fe$_{1-x}$Co$_x$)$_2$As$_2$ and Sr(Fe$_{1-x}$Co$_x$)$_2$As$_2$ Measured with Muon Spin Relaxation}

\author{T.~J.~Williams}
 \affiliation{Department of Physics and Astronomy,
   McMaster University, 1280 Main St. W.,
   Hamilton, ON, Canada, L8S 4M1}

\author{A.A. Aczel}
 \affiliation{Department of Physics and Astronomy,
   McMaster University, 1280 Main St. W.,
   Hamilton, ON, Canada, L8S 4M1}

 \author{E. Baggio-Saitovitch}
 \affiliation{Centro Brasilieiro de Pesquisas Fisicas, Rua Xavier Sigaud 150 Urca, CEP 22290-180 Rio de Janeiro, Brazil}

     \author{S.~L.~Bud'ko}
     \affiliation{Department of Physics and Astronomy and Ames Laboratory, Iowa State University, Ames, Iowa 50011, USA}

     \author{P.~C.~Canfield}
     \affiliation{Department of Physics and Astronomy and Ames Laboratory, Iowa State University, Ames, Iowa 50011, USA}

\author{J.P. Carlo}
 \affiliation{Department of Physics,
   Columbia University,
   538 W. 120th St., New York, NY, 10027}

  \author{T. Goko}
     \affiliation{Department of Physics and Astronomy,
   McMaster University, 1280 Main St. W.,
   Hamilton, ON, Canada, L8S 4M1}
     \affiliation{Department of Physics,
   Columbia University,
   538 W. 120th St., New York, NY, 10027}
     \affiliation{TRIUMF, Vancouver, British Columbia, Canada, V6T 2A3}
   
 \author{H. Kageyama}
 \affiliation{Department of Chemistry, Kyoto University, Kyoto 606-8502, Japan}

 \author{A. Kitada}
 \affiliation{Department of Chemistry, Kyoto University, Kyoto 606-8502, Japan}

 \author{J. Munevar}
 \affiliation{Centro Brasilieiro de Pesquisas Fisicas, Rua Xavier Sigaud 150 Urca, CEP 22290-180 Rio de Janeiro, Brazil}

     \author{N.~Ni}
     \affiliation{Department of Physics and Astronomy and Ames Laboratory, Iowa State University, Ames, Iowa 50011, USA}

 \author{S. R. Saha}
 \affiliation{Center for Nanophysics and Advanced Materials,
   Department of Physics, University of Maryland, College Park, MD
   20742, USA}

\author{K.~Kirshenbaum}
\affiliation{Center for Nanophysics and Advanced Materials, Department of Physics, University of Maryland, College Park, MD 20742, USA}

 \author{J. Paglione}
 \affiliation{Center for Nanophysics and Advanced Materials, Department of Physics, University of Maryland, College Park, MD 20742, USA}

 \author{D.R. Sanchez-Candela}
 \affiliation{Centro Brasilieiro de Pesquisas Fisicas, Rua Xavier Sigaud 150 Urca, CEP 22290-180 Rio de Janeiro, Brazil}

 \author{Y.J. Uemura}
 \affiliation{Department of Physics,
   Columbia University,
   538 W. 120th St., New York, NY, 10027}

 \author{G.M. Luke}
      \altaffiliation{author to whom correspondences should be addressed: E-mail:[luke@mcmaster.ca]}
\affiliation{Department of Physics and Astronomy,
   McMaster University, 1280 Main St. W.,
   Hamilton, ON, Canada, L8S 4M1}
    \affiliation{Canadian Institute of Advanced Research, Toronto, Ontario, Canada, M5G 1Z8}

\date{\today}

\begin{abstract}
We report muon spin rotation ($\mu$SR) measurements of single crystal Ba(Fe$_{1-x}$Co$_x$)$_2$As$_2$ and Sr(Fe$_{1-x}$Co$_x$)$_2$As$_2$. From measurements of the magnetic field penetration depth
$\lambda$ we find that for optimally- and over-doped samples, $1/\lambda(T\rightarrow0)^2$ varies monotonically with the superconducting transition temperature T$_{\rm C}$. Within the superconducting state we observe a positive shift in the muon precession signal, likely indicating that the applied field induces an internal magnetic field. The size of the induced field decreases with increasing doping but is present for all Co concentrations studied.
\end{abstract}

\pacs{76.75.+i}

\maketitle
\section{Introduction}
Of the various families of iron pnictide superconductors, the
so-called 122 family has been extensively studied due to their high
$T_C$'s and the ability to grow single crystals. This family includes
BaFe$_2$As$_2$ and SrFe$_2$As$_2$. Unlike the cuprates, these
materials are quite robust against in-plane disorder, brought about by
electron-doping for Fe atoms either by Co, Ni or other transition metals. The transition
temperatures remain fairly high for these substitutions, with $T_C$ =
22K for Ba(Fe$_{0.926}$Co$_{0.074}$)$_2$As$_2$\cite{Sefat_08c,Ni_08},
20.5K for Ba(Fe$_{0.952}$Ni$_{0.048}$)$_2$As$_2$\cite{Canfield_09}, 23K for
Ba(Fe$_{0.9}$Pt$_{0.1}$)$_2$As$_2$\cite{Saha_10}, 14K for
Ba(Fe$_{0.961}$Rh$_{0.039}$)$_2$As$_2$\cite{Ni_09}, 19.5K for Sr(Fe$_{0.8}$Co$_{0.2}$)$_2$As$_2$\cite{Kim_09} and 9.5K for Sr(Fe$_{0.925}$Ni$_{0.075}$)$_2$As$_2$\cite{Saha_09}.

Measurements of the penetration depth and superfluid density have attempted to address the nature of the superconducting gap symmetry. NMR has shown the lack of a coherence peak\cite{Nakai_08,Grafe_08}, indicative of unconventional pairing. Similarly, tunnel-diode resonator measurements in Ba(Fe$_{1-x}$Co$_x$)$_2$As$_2$ also show power-law temperature dependences for the penetration depth\cite{Gordon_09a,Gordon_09b}, which are interpreted in terms of gap nodes. Other measurements see a constant superfluid density at low temperatures\cite{Martin_09,Leutkens_08}, indicating an {\it s}-wave gap. Likewise, the possibility of multi-band superconductivity in the pnictides has also been studied by analyzing the temperature-dependence of the superfluid density\cite{Williams_09,Terashima_09}.

\section{Experimental}
Muon spin rotation ($\mu$SR) is a powerful local microscopic tool for characterizing the magnetic properties of materials, in superconducting or other states. A thorough description of the application of $\mu$SR to studies of superconductivity can be found elsewhere\cite{Sonier_07}. In a transverse field (TF) $\mu$SR experiment, spin-polarized positive muons are
implanted one at a time into a sample.
 Each muon spin precesses around the local magnetic field until the muon decays into a positron, which is preferentially ejected along the direction of the muon spin at the time of decay (as well as two neutrinos which are not detected). In the presence of a vortex lattice, the spatial variation of the magnetic field distribution results in a dephasing of the muon spin polarization and a relaxation of the precession signal. A Fourier transform of the spin polarization function essentially reveals the field distribution which exhibits a characteristic Abrikosov lineshape. The lineshape (or equivalently the relaxation function in the time domain) depends on the lattice geometry, magnetic field penetration depth $\lambda$, coherence length, $\xi$, and the amount of lattice disorder. As a result, careful analysis of the relaxation
 function allows these microscopic parameters to be determined in the vortex state. Such measurements demonstrated the presence of gap nodes characteristic of d-wave pairing
 in high quality single crystals of YBa$_2$Cu$_3$O$_{6.97}$\cite{Sonier_07}. In ceramic samples this anisotropic lineshape is generally not
 observed, rather the broadened line is generally well described by a Gaussian distribution; however, the width of this distribution (the Gaussian relaxation rate) $\sigma$ has been shown to be
 proportional to the superfluid density divided by the effective mass $\sigma\propto n_s/m^*\propto 1/\lambda^2$\cite{Brandt_88,Maisuradze_09}. Previous studies of cuprates found that extrinsic effects in ceramics can result in
 the correct temperature dependence of the superfluid density being masked; for this reason, reliable measurements of the superfluid density require the use of single crystals and the observation of
 an anisotropic lineshape characteristic of a vortex lattice.

High quality single crystals of Ba(Fe$_{1-x}$Co$_{x})_2$As$_2$ with x=0.061, 0.074, 0.107 and 0.114 were grown at Ames from self flux as described in detail elsewhere\cite{Ni_08}. Some measurements of the $x=0.074$ sample were reported previously\cite{Williams_09}. A single crystal of Sr(Fe$_{1-x}$Co$_{x})_2$As$_2$ with $x=0.13$ was grown at Maryland, also from self flux\cite{Saha_09b}.
The crystals, each of roughly 1~cm$^2$ area, were mounted in a helium gas flow cryostat on the M20 surface muon beamline at TRIUMF, using a low background arrangement such that only positrons originating from muons landing in the specimens were
collected in the experimental spectra. Zero field $\mu$SR measurements of each sample confirmed that no magnetic order or spin freezing was present in any of the samples.

\section{Penetration Depth}
\begin{figure}[htb]
\includegraphics[angle=0,width=\columnwidth]{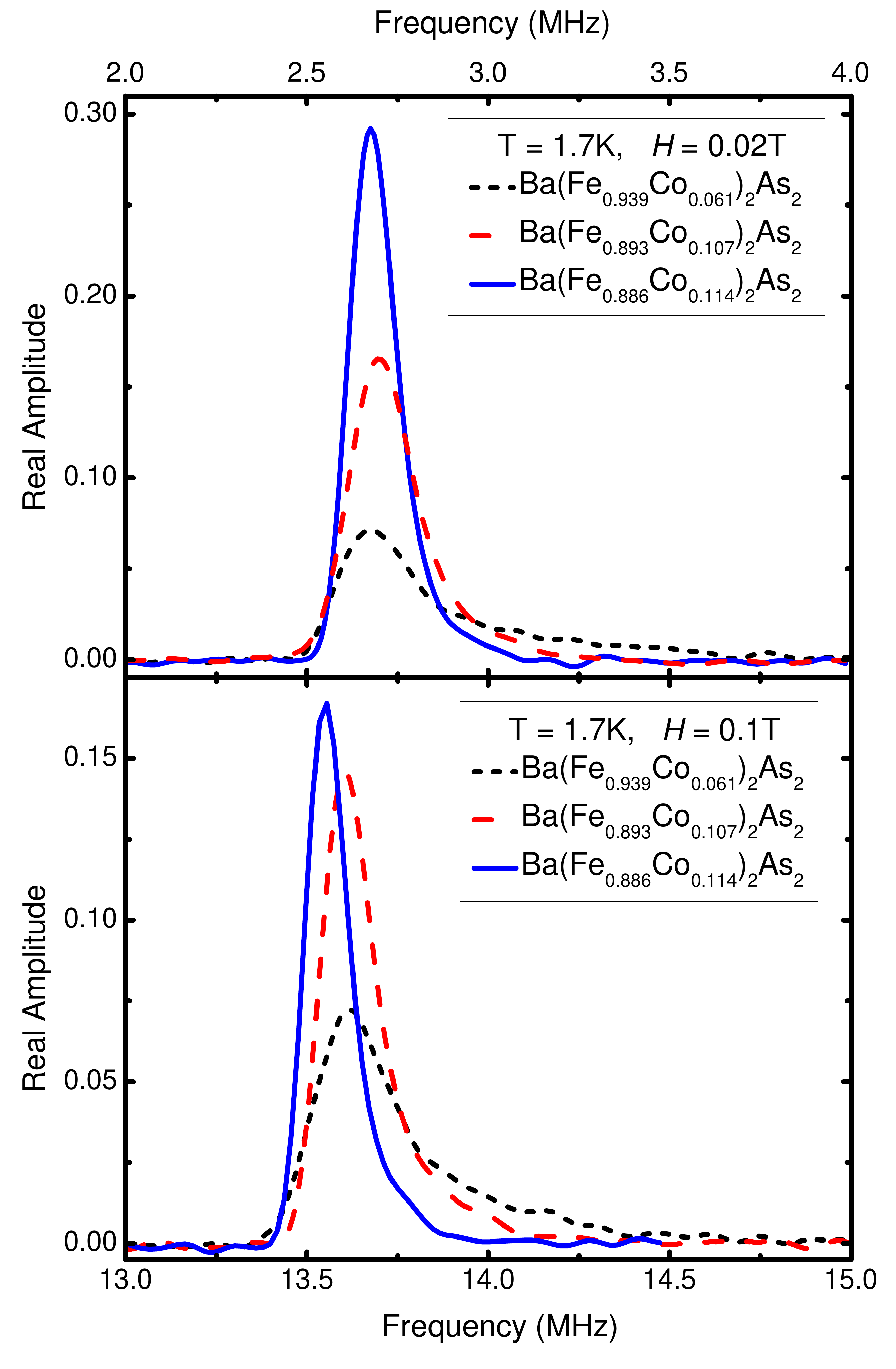}
\caption{\label{fft_data}
(Color online)
Fourier transforms of TF-$\mu$SR spectra for Ba(Fe$_{1-x}$Co$_x$)$_2$As$_2$, showing anisotropic lineshapes characteristic of an Abrikosov vortex lattice.}
\end{figure}

Fourier transforms of the transverse field (TF)-$\mu$SR spectra (a representative set are shown in Fig.~\ref{fft_data}) exhibit the anisotropic lineshape characteristic of an Abrikosov lattice, indicating the presence of a least locally well-ordered vortices
in the superconducting state.   All of the  the Fourier transform lineshapes are consistent with a triangular vortex lattice; for example, a square lattice would have the frequency corresponding to the most likely field (the peak of the lineshape) much more  separated from the minimum field in the field distribution.
We analyzed the data by fitting the spectra to an analytical Ginzburg-Landau model which allows us to calculate theoretical $\mu$SR time spectra as a function of the
vortex lattice geometry, magnetic field penetration depth ($\lambda$) and coherence length ($\xi$). We included the effects of vortex lattice disorder in our analysis via an additional Gaussian broadening of our $\mu$SR spectrum\cite{Brandt_88, Riseman_95}, where we assumed that this broadening was proportional to $1/\lambda^2$ as observed in previous studies of cuprates and other high $\kappa$ superconductors~\cite{Sonier_07}. The errors quoted in various fit parameters included the correlations between the various parameters. The fit parameters were fairly weakly correlated since the effect of each parameter on the relaxation function is reasonably unique: the penetration depth affects the overall linewidth, the coherence length affects the high field cutoff while disorder gives an overall broadening of the various van Hove singularities in the lineshape.  Consistent with our previous measurements of Ba(Fe$_{0.926}$Co$_{0.074})_2$As$_2$ \cite{Williams_09}, we found that the rms deviation of the vortex positions
$(\langle s^2\rangle^{1/2})$ relative to the vortex separation was greatest in lower fields (up to 30\% in 0.02~T at low temperature) and smallest at the highest fields (about 2\% in 0.1~T) and decreased with increasing
temperature. The disorder was greatest for the samples with the highest T$_c$.

\begin{figure}[htb]
\includegraphics[angle=0,width=\columnwidth]{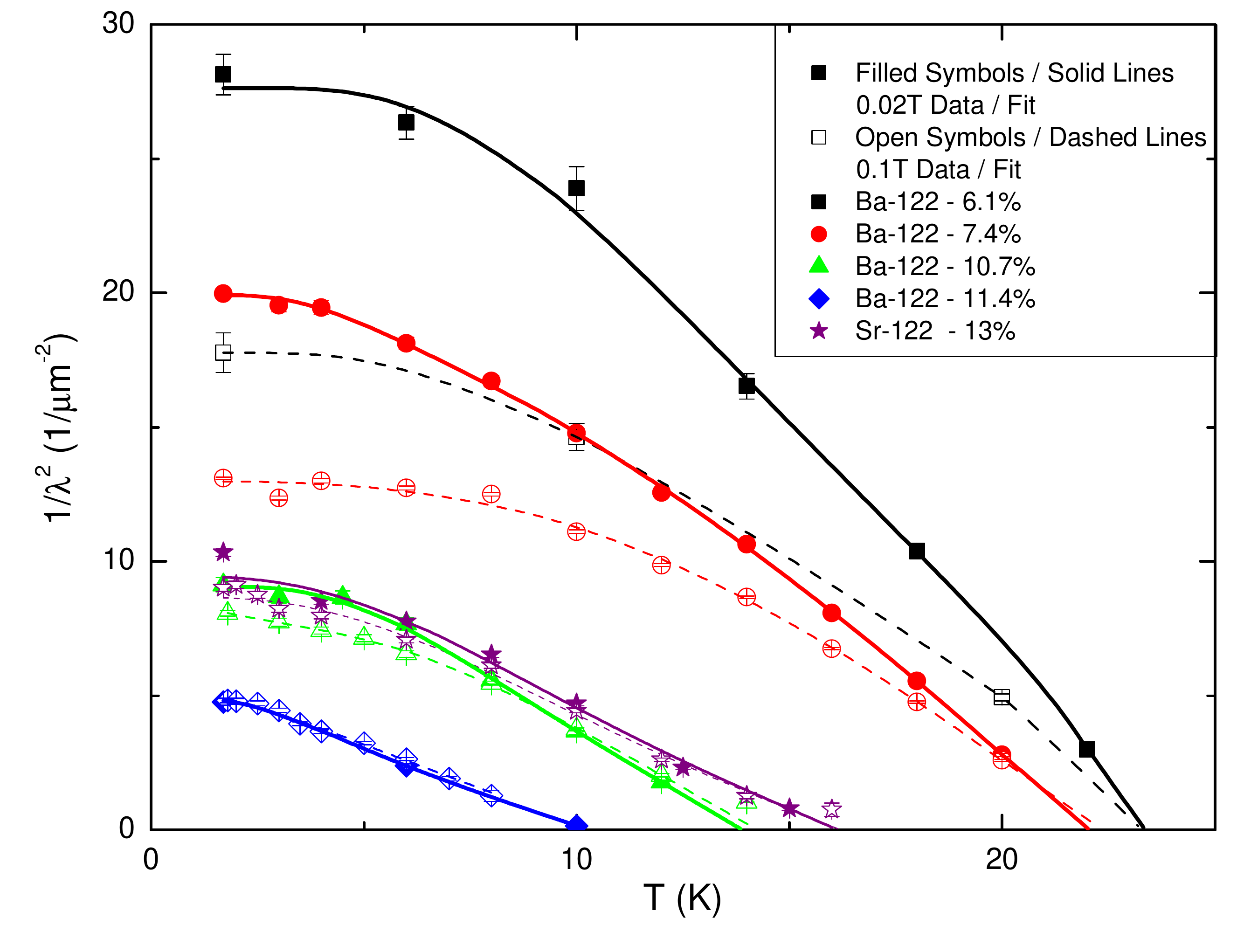}
\caption{\label{superfluid}
(Color online)
Measured $1/\lambda^2$ for Ba(Fe$_{1-x}$Co$_x$)$_2$As$_2$ and Sr(Fe$_{1-x}$Co$_x$)$_2$As$_2$ measured in TF=0.02~T (filled symbols) and 0.1~T (open symbols).}
\end{figure}

Results of this analysis for $1/\lambda^2$
are shown in Fig.~\ref{superfluid} for applied fields of 0.1~T, and
0.02~T. In conventional weak-coupling BCS theory, the low-temperature
behaviour of $1/\lambda^2$ should be exponentially flat while the presence of gap
nodes would be reflected in low-temperature power law behaviour. We see in Fig.~\ref{superfluid} that
 the low-temperature behaviour varies more rapidly than standard
BCS predictions, and also note that recent specific heat measurements have observed
the possibility of gap nodes\cite{Reid_10}. Following our earlier work on the Ba(Fe$_{0.926}$Co$_{0.074})_2$As$_2$\cite{Williams_09} we fit the superfluid density to a phenomenological two-gap model\cite{Bouquet_01b,Ohishi_03} which has been employed in previous $\mu$SR studies of LaFeAs(O,F), Ca(Fe,Co)AsO, and (Ba,K)Fe$_2$As$_2$\cite{Takeshita_08},
\begin{equation}n_s(T)=
n_s(0)-w\cdot\delta n_s(\Delta_1,T)-(1-w)\cdot\delta n_s(\Delta_2,T)
\label{eq1}
\end{equation}
where $w$ is the relative weight for the first gap, $\Delta_1$. Here, the gap functions are given by,
\begin{equation}\delta n(\Delta,T)=
\frac{2n_s(0)} {k_BT}\int_{0}^{\infty}f(\epsilon,T)\cdot[1-f(\epsilon,T)]d\epsilon
\label{eq2}
\end{equation}
where $f(\epsilon,T)$ is the Fermi distribution given by,
\begin{equation}f(\epsilon,T)=
(1+e^{\sqrt{\epsilon^2+\Delta(T)^2}/k_BT})^{-1}
\label{eq3}
\end{equation}
Here, $\Delta_i$ ($i=1$ and $2$) are the energy gaps at $T=0$, and $\Delta_i(T)$ were taken to follow the standard weak-coupled BCS temperature dependence.
This model reduces to a single-gap BCS model when $w=1$.
The size of the gaps, $\Delta_1$ and $\Delta_2$, and $T_{\rm C}$ were fit globally, while $n_s(0)$ and the weighting factor, $w$, were allowed to be field-dependent.
The fit values of $1/\lambda^2$ are shown by the fit lines on Fig.~\ref{superfluid}. We see that this two gap model fits the observed
temperature dependence of $1/\lambda^2$ for each of the samples and fields measured. Single gap fits did not give satisfactory results (when the BCS gap function was used).
For most of the samples we obtained the larger gap value $2\Delta/k_BT_c\approx 3.7$ which is close to the
weak coupled BCS value of 3.5. For the Sr(Fe$_{0.87}$Co$_{
  0.13}$)$_2$As$_2$ the larger gap was $2\Delta/k_BT_c\approx 2.7$,
less than the BCS value. For the Ba(Fe$_{1-x}$Co$_x$)$_2$As$_2$ samples with
$x=0.107,\, 0.114$ most of the weight was on the smaller gap. The results for these three samples give a stronger low temperature dependence to the superfluid density
than for a single-gap weak-coupled BCS system, as can be seen on Fig.~\ref{superfluid}. This steeper temperature dependence may possibly reflect a non-s-wave gap as has been interpreted by
tunnel diode resonator measurements.\cite{Gordon_09a, Gordon_09b} We don't have enough data points, especially at temperatures below 2~K, to make a definitive statement regarding the presence of gap nodes.
We are able, however, to reliably extrapolate the superfluid density to obtain a good estimate of the magnitude of $1/\lambda^2(T\rightarrow0)$. Examining the behaviour of $1/\lambda^2$ for the different samples in Fig.~\ref{superfluid} we see that there is considerable variation in
$1/\lambda^2({\rm T}\rightarrow0)$.
Over the range of dopings and fields studied, the value of $1/\lambda^2(T\rightarrow0)$ varies from 5 $\mu m^{-2}$ to nearly 30 $\mu m^{-2}$, more than half an order of magnitude.

There is considerable field dependence in $1/\lambda^2({\rm T}\rightarrow0)$ for the Ba(Fe$_{1-x}$Co$_x$)$_2$As$_2$ samples with $x=0.061$ and $x=0.074$  which is essentially absent for the higher doped samples with the smaller superfluid density. We first noted this large field dependence in our study of Ba(Fe$_{0.926}$Co$_{0.074}$)$_2$As$_2$\cite{Williams_09}; subsequent studies on other
pnictides have seen similar behaviour.\cite{Weyeneth_09} The field dependence in the density of states in a multiband superconductor has been calculated by Ichioka {\em et al.}\cite{ichioka_04} who noted that a strong field dependence is expected for fields on the order of the smaller gap size.  Results of previous $\mu$SR measurements of a variety of  multi-gap superconductors are described in Ref.~\cite{Sonier_07}.  In those materials (such as NbSe$_2$) the origin of the field dependence is the loosely bound core states associated with the smaller gap.  With increasing field these states become more delocalized and affect the field distribution seen by the muon ensemble. The field dependence we observe in this study is larger than we would expect to be due to the smaller gap $\Delta_2$ and may possibly indicate that an anisotropic gap (perhaps with nodes) might be more appropriate for the smaller gap than the uniform gap model used to fit the temperature dependence of the superfluid density. We note that different gap symmetries on different parts of the Fermi surface might resolve the apparent discrepancies between different techniques that probe the normal state carrier concentration (eg. tunnel diode oscillator, microwave) and superfluid carrier concentrations ($\mu$SR).  To estimate the zero applied field values of $\lambda_0$ we have performed a linear extrapolation of the fit values 
of $1/\lambda_0^2$ measured in 0.02~T and 0.1~T and included the resulting $\lambda_0(B\rightarrow0)$ values in Table~\ref{props}.

\begin{table}[tbp]
\begin{tabular}{|l|l|l|l|l|}
\hline
 & ${\rm T_{C}}$&$\lambda_0 (0.02T)$ &$\lambda_0( 0.1T)$ &$\lambda_0({\rm B}=0)$
  \\
\hline
Ba(Fe$_{0.939}$Co$_{0.061}$)$_2$As$_2$& 23.6& $189.4\pm1.1$ & $240.5\pm2.0$&$182.6\pm1.4$\\
Ba(Fe$_{0.926}$Co$_{0.074}$)$_2$As$_2$& 22.1 & $224.2\pm0.6$ &$277.4\pm1.0$&$216.8\pm0.7$\\
Ba(Fe$_{0.899}$Co$_{0.101}$)$_2$As$_2$& 14.1 & $332.2\pm2.2$& $348.3\pm4.6$&$329.3\pm3.4$\\
Ba(Fe$_{0.89}$Co$_{0.11}$)$_2$As$_2$& 10.3 & $453.8\pm 2.6$&$448.0\pm 2.4$&$454.9\pm3.6$ \\
Sr(Fe$_{0.87}$Co$_{0.13}$)$_2$As$_2$& 16.2 & $325.5\pm 0.5$ &$ 339.8\pm 0.6$&$322.8\pm0.7$\\
\hline
\end{tabular}
\caption[]{Results of fitting $1/\lambda^2(T)$ to Eqn.~\ref{eq1} for T$_c$(K), $\lambda_0$ (nm) in fields of 0.02~T and 0.1~T.  Also shown are values of $\lambda_0$(nm) extrapolated to zero field.}
\label{props}
\end{table}

\begin{figure}[htb]
\includegraphics[angle=0,width=\columnwidth]{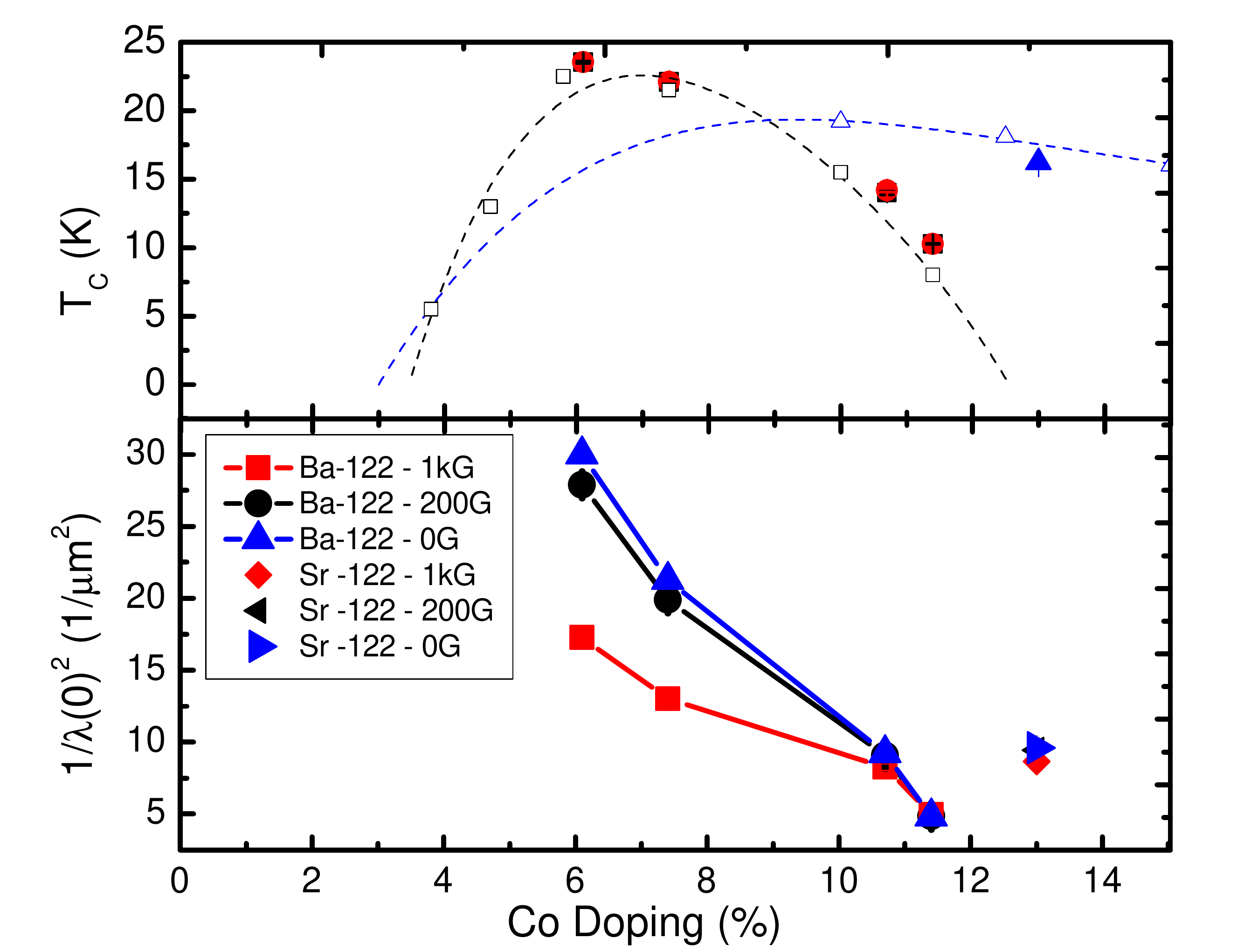}
\caption{\label{doping}
(Color online) Superconducting Tc's and $1/\lambda^2({\rm T}\rightarrow0)$ for Ba(Fe$_{1-x}$Co$_x$)$_2$As$_2$ and Sr(Fe$_{1-x}$Co$_x$)$_2$As$_2$ as a function of Co concentration $x$,
measured in TF=0.02T and 0.1T, and extrapolated to ${\rm B}=0$. The open points and dashed lines are
the measured T$_C$'s and the superconducting dome taken from
\cite{Ni_08} for Ba(Fe,Co)$_2$As$_2$ and \cite{Leithe-Jasper_08,Saha_09} for
Sr(Fe,Co)$_2$As$_2$. }
\end{figure}

 Fig.~\ref{doping} shows the extrapolated values of $1/\lambda^2({\rm T}\rightarrow0)$ and the fit values of T$_{\rm C}$ as a function of the
level of Co doping $x$ for Ba(Fe$_{1-x}$Co$_x$)$_2$As$_2$ and Sr(Fe$_{1-x}$Co$_x$)$_2$As$_2$. We see that above $x=0.06$ in Ba(Fe$_{1-x}$Co$_x$)$_2$As$_2$ the superconducting T$_c$ decreases with increasing
Co substitution, in agreement with previous work. Additionally, the T$_c$ for Sr(Fe$_{1-x}$Co$_x$)$_2$As$_2$ does not lie on the same curve as for the Ba(Fe$_{1-x}$Co$_x$)$_2$As$_2$ family; the location of the superconducting phase dome within the phase diagram is different for the two families. The lower panel of Fig.~\ref{doping} shows the evolution of the extrapolated $1/\lambda^2({\rm T}\rightarrow0)$. Within the Ba(Fe$_{1-x}$Co$_x$)$_2$As$_2$ family, there is a monotonic decrease in $1/\lambda^2({\rm T}\rightarrow0)$ and again,
 the point for Sr(Fe$_{1-x}$Co$_x$)$_2$As$_2$ does not lie on the same curve.

\begin{figure}[htb]
\includegraphics[angle=0,width=\columnwidth]{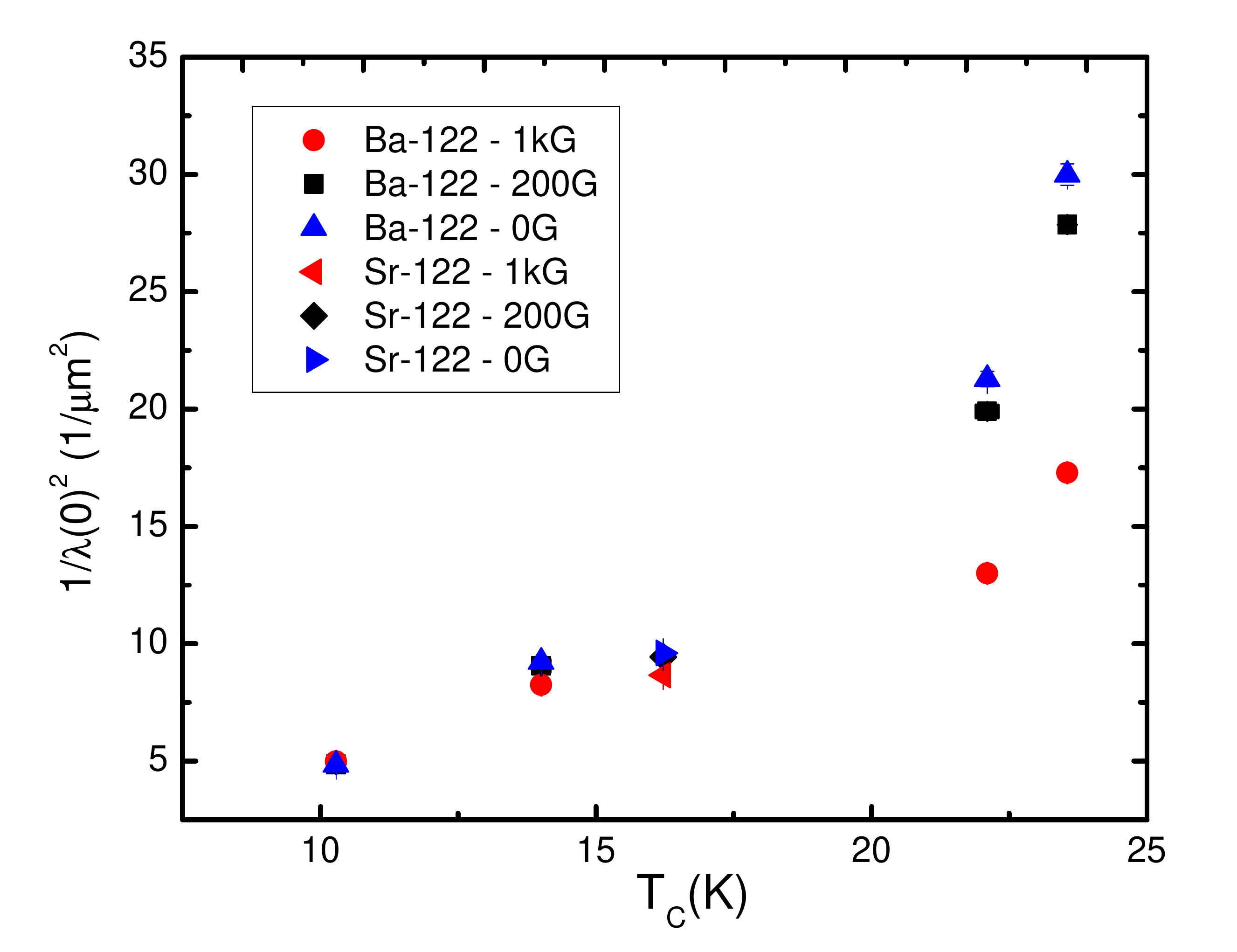}
\caption{\label{correlate}
(Color online) Superconducting T$_{\rm C}$ vs. $1/\lambda^2({\rm T}\rightarrow0$) for Ba(Fe$_{1-x}$Co$_x$)$_2$As$_2$ and Sr(Fe$_{1-x}$Co$_x$)$_2$As$_2$ as a function of Co concentration $x$,
measured in TF=0.02T and 0.1T and extrapolated to ${\rm B}=0$. }
\end{figure}

Muon spin rotation measurements on a wide variety of cuprate and other exotic superconductors have revealed a strong, roughly linear correlation between the superconducting transition temperature and
the extrapolated zero temperature superfluid density divided by the effective mass.\cite{uemura_89} This relation is not expected in standard BCS theory, implying that a different mechanism is responsible for superconductivity in these systems. We plot our fit values of T$_{\rm C}$ vs. $1/\lambda^2({\rm T}\rightarrow0)$ in Fig.\ref{correlate} for Ba(Fe$_{1-x}$Co$_x$)$_2$As$_2$ and Sr(Fe$_{0.87}$Co$_{0.13}$As$_2$)$_2$.
In contrast to the plots on Fig.~\ref{doping}, the points for all of the samples lie close to common curves for both 20~mT and 100~mT. We see that in single crystals of Ba(Fe$_{1-x}$Co$_x$)$_2$As$_2$ and Sr(Fe$_{1-x}$Co$_x$)$_2$As$_2$, the superconducting T$_{\rm C}$ is apparently determined by the carrier density divided by the effective mass.

\begin{figure}[htb]
\includegraphics[angle=0,width=\columnwidth]{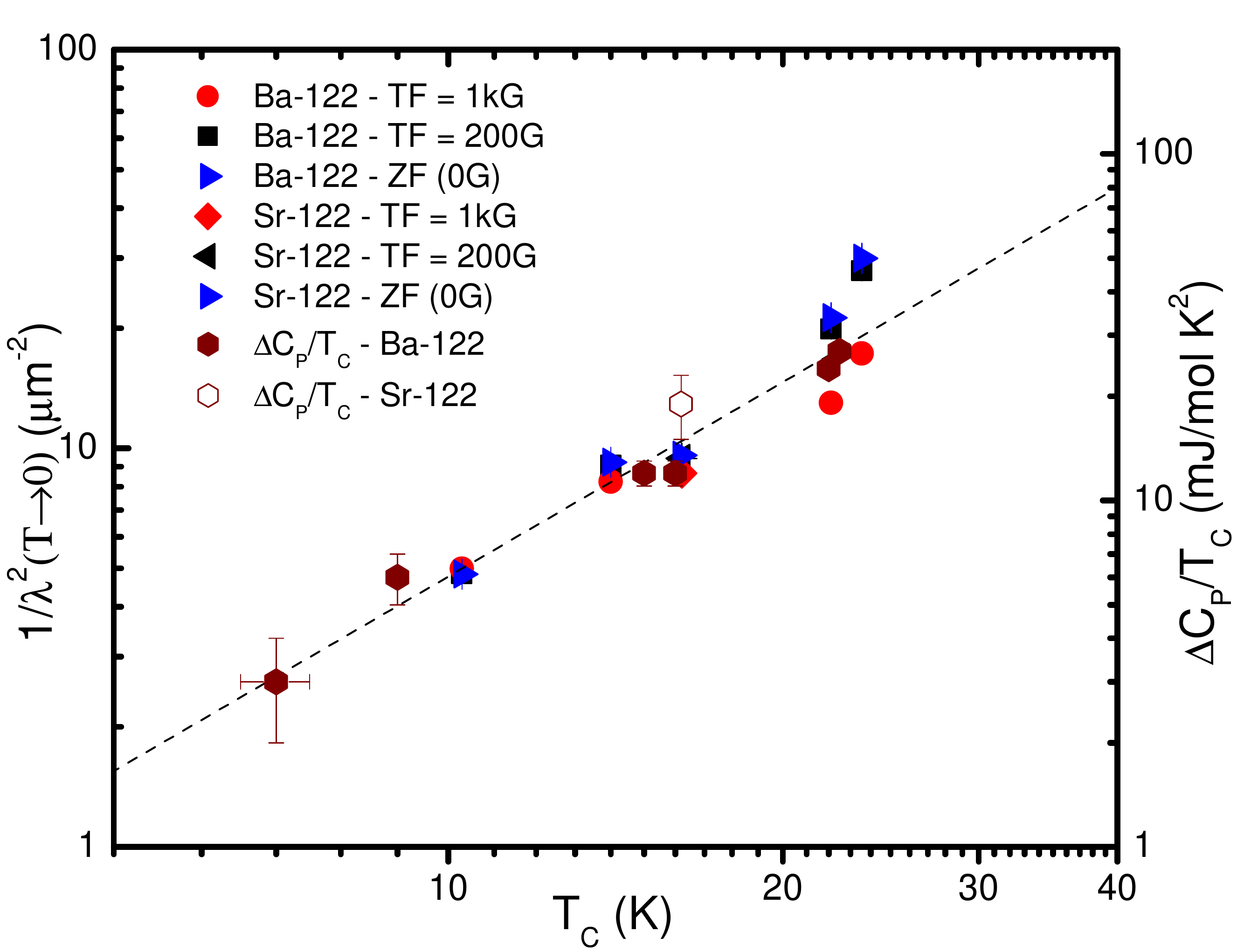}
\caption{\label{correlate_log}
(Color online) $1/\lambda^2({\rm T}\rightarrow0$) vs. superconducting T$_{\rm C}$ for Ba(Fe$_{1-x}$Co$_x$)$_2$As$_2$ and Sr(Fe$_{1-x}$Co$_x$)$_2$As$_2$ as a function of Co concentration $x$,
measured in TF=0.02T and 0.1T and extrapolated to ${\rm B}=0$, plotted with $\Delta$C$_P$/T$_c$ from Ref.~\cite{budko_09}. The dashed line has a slope $n=2$ and is a guide to the eye.}
\end{figure}

Specific heat measurements\cite{budko_09} of the superconducting transition found that the specific heat jump at T$_c$ divided by T$_c$ was correlated with T$_c$ as
$\Delta C_P/T_c\propto T_c^2$. Our results for $1/\lambda(T\rightarrow0)^2$ are shown, along with specific heat jump for Sr(Fe$_{0.87}$Co$_{0.13}$)$_2$As$_2$ and
 the results of Bud'ko {\em et al.}\cite{budko_09} in Fig.~\ref{correlate_log}. In agreement with the specific heat, we find that $1/\lambda(T\rightarrow0)^2$ can be well described by a straight line with slope $n\approx2$ as indicated by the dashed line. The common variation of the
superfluid density\cite{uemura_93} and the specific heat jump $\Delta C_P/T_c$ and T$_c$,\cite{loram_94} as a function of carrier doping was
first noted in overdoped Tl$_2$Ba$_2$CuO$_{6+\delta}$ cuprates\cite{uemura_04}. The present case of the FeAs superconductors, shown in Fig.~\ref{correlate_log}, exhibits
commonalities to the cuprates in this regard.

There are two contributions to the measured penetration depth with chemical substitution. First, doping changes the carrier concentration which directly changes the London penetration depth $\lambda_L$
via $1/\lambda_L^2\propto n_s/m^*$. If a system is in the dirty limit, the measured penetration depth is actually an effective penetration depth $\lambda_{eff}=\lambda_L(1+\xi_0/l)^{\frac{1}{2}}$ where $\xi_0$ is the coherence length and
$l$ is the mean free path.\cite{tinkham} The upper critical field is quite large in these systems, taking H$_{c2}\approx 50$~T gives an estimate of $\xi_0=2.5$~nm. In order to estimate the mean free path, a reasonable value of the Fermi velocity is needed. Due
to the unclear situation of the nature of the Fermi surface in the pnictides, an estimate of the pair-breaking effect is unlikely to be accurate. Optical conductivity measurements have directly detected the opening of the superconducting gap in Ba(Fe$_{1-x}$Co$_x$)$_2$As$_2$ with $x=0.1$\cite{gorshunov_09}, $x=0.07$\cite{heumen_09} and $x=0.065$\cite{kwkim_09}. Broadening of the normal-state zero frequency Drude conductivity indicates the presence of significant normal-state scattering. Some authors have argued that strong pair-breaking can account for a number of effects, including the specific heat jump at T$_C$ and the behaviour of
$dH_{c2}/dT$\cite{Gordon_09}.  Assuming strong pair-breaking, Kogan\cite{vgkogan_10} has found that $1/\lambda(0)\propto{\rm T}_c$, in agreement with our results in Fig.~\ref{correlate_log}.
However, the existence of such strong pair breaking is not yet proven. Although scattering is clearly present in these systems, it is unlikely that reasonably modest changes in the dopant concentration (of a few percent) could cause such a dramatic change in the scattering so as to dominate the penetration depth and as such, substantial changes in the superfluid density are apparently occurring with chemical substitution.

Hall effect measurements show that the normal-state carrier
concentration increases monotonically with increased chemical
substitution\cite{Mun_09}. If in fact the superfluid density decreases
with increasing doping above the maximum T$_c$, then this implies that
not all of the carriers join the condensate below T$_c$. This
segregation into superconducting and normal fluids could be in
reciprocal space, if superconductivity occurs on only some parts
of the Fermi surface. This could also occur in real space with phase separation into normal and superconducting regions. Previous $\mu$SR measurements of overdoped
Tl$_2$Ba$_2$CuO$_{6+\delta}$\cite{uemura_93, niedermayer_93} exhibited similar behaviour, with increased normal state doping and a loss of superconducting carrier density.
Real space phase separation has been seen in other $\mu$SR measurements of both hole-\cite{Takeshita_09,Aczel_08} and electron-doped\cite{Goko_09} pnictide superconductors.

Phase separation (either in real space or reciprocal space) should leave a residual normal fluid whose spectral weight should increase with Co substitution and which should be apparent in
measurements of optical conductivity. Recent optical measurements by Gorshunov {\em et al.}, in Ba(Fe$_{1-x}$Co$_x$)$_2$As$_2$ with $x=0.1$ do in fact show appreciable residual conductivity
well within the superconducting state which may be evidence of this residual normal fluid\cite{gorshunov_09}. Additional measurements at low frequencies for a range of doping are needed to further test this hypothesis. Gofryk {\em et al.} reported specific heat measurements of
Ba(Fe$_{1-x}$Co$_x$)$_2$As$_2$ for a range of Co concentrations.\cite{gofryk_10}. They found a substantial normal fluid response (residual linear specific heat contribution) which increased
with Co concentration for $x>0.08$.

Our $\mu$SR results in Co substituted BaFe$_2$As$_2$ and SrFe$_2$As$_2$
indicate that the vortex lattice
exists throughout the samples which indicates that any phase separation is either in real space with a characteristic lengthscale much less than the penetration depth (perhaps the coherence length) or in
reciprocal space. A model for real space phase separation for overdoped cuprates has been discussed in Ref.~\cite{uemura_01} . If the phase separation occurs in momentum space, it could originate
perhaps from only some of the multiple bands in these systems participating in the pairing. Angle-resolved photoemission measurements\cite{sekiba_09} have shown that above $x\approx 0.08$ overdoped electrons fill the hole Fermi surface at the Brillouin zone centre, resulting in a loss of interband scattering. If this scattering is
responsible for pairing, then the loss of the hole states with substitution could reduce the superfluid density, even though the normal state carrier concentration increases with doping.

\section{Paramagnetic Frequency Shift}
\begin{figure}[htb]
\includegraphics[angle=0,width=\columnwidth]{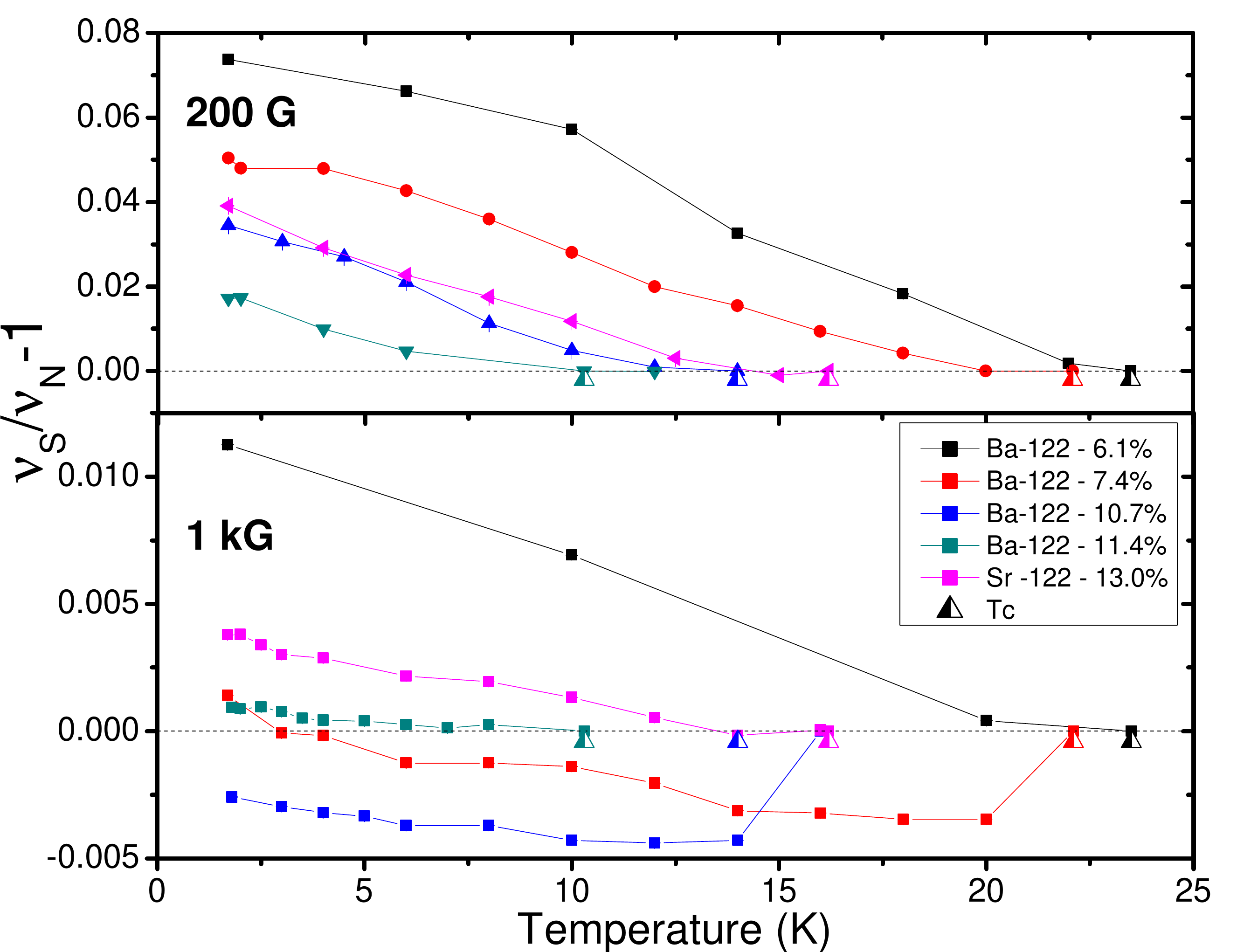}
\caption{\label{frq}
(Color online) Fractional shift of muon precession frequency $\nu_{\mu}/\nu_{\mu N} -1$ relative to the normal state frequency $\nu_{\mu N}$. Superconducting transition temperatures indicated by triangle symbols. }
\end{figure}

 When fitting the $\mu$SR time spectra to our analytical Ginzburg-Landau model, one of the fitted parameters is the average muon precession frequency $\nu_\mu$. In the normal state, this precession frequency is given by $\nu_\mu=(1+K_\mu)\gamma_\mu{\rm B}_{\rm ext}$ where B$_{\rm ext}$ is the externally applied magnetic field, $\gamma_\mu$ is the muon gyromagnetic ratio and K$_\mu$ is the Knight shift. In the superconducting state the muon precession frequency is generally slightly reduced from the normal state value due to flux expulsion; for thin plate-like samples this reduction is generally quite small due to the demagnetizing factor. We show the fitted values of the fractional shift in the precession frequency relative to its normal state value $\nu_{\mu N}$ in Fig.~\ref{frq}. We see that except for a negative shift right below T$_C$ for some samples due to bulk screening, all samples have increasing frequency shifts with decreasing temperature in the superconducting state.  A similar positive frequency shift has also been reported by Khasanov {\em et al.}  in SrFe$_{1.75}$Co$_{0.25}$As$_2$\cite{khasanov_prl_103}.   A positive value of
 $\nu_{\mu}/\nu_{\mu N} -1$ indicates that the field at the muon site is actually {\em greater} than the applied field. Since bulk screening can only contribute a negative frequency shift, we need to find a different explanation for our observed positive shifts.

 The fractional shift within the superconducting state is considerably larger in the 0.02T data than in the 0.1T runs as shown in Fig.~\ref{frq}. In fact, the absolute value of the shifts ($\nu_\mu-\nu_{\mu N}$) is roughly the same for the two fields. The shifts are also largest for the samples with the highest T$_C$ and highest superfluid density $n_s/m^*\propto 1/\lambda^2$. Previous $\mu$SR studies of the
 electron-doped cuprate superconductor Pr$_{2-x}$Ce$_x$CuO$_4$ also exhibited a positive frequency shift below T$_C$ which was interpreted as evidence of field-induced magnetism\cite{Sonier_03}. In that case, the absolute shift decreased with increasing field (not just the fractional shift), indicating that the induced fields were perpendicular to the applied field. In the present case, the fact that the absolute shift is roughly field-independent indicates that the induced moments must be parallel to the applied field and have a ferromagnetic character (antiferromagnetic fields would split the precession line, rather than shift it). We note that these field-induced ferromagnetic fields would not be apparent in bulk susceptibility measurements, since they would be screened by supercurrents on the surface of the sample.
 In each sample the paramagnetic frequency shift sets in at the superconducting T$_c$ of each particular sample, implying that it is a property of the superconducting state. One possible source of such a
 field could be a spin triplet pair state where the Cooper pairs possess a non-zero angular momentum. However, other explanations are also possible and further experiments will be required to
 determine the microscopic origin of these internal fields.

\section{Conclusions}
We have measured the London penetration depth in single crystals of Ba(Fe$_{1-x}$Co$_x$)$_2$As$_2$ (with $x\geq0.061$)and Sr(Fe$_{0.87}$Co$_{0.13}$)$_2$As$_2$ using muon spin rotation.
The temperature dependence of $1/\lambda^2$ can be fit by a two-band model where the gaps follow the BCS temperature dependence. In the more highly-doped samples we find that
dominant gap magnitude is considerably smaller than the weak-coupled BCS result, implying that the gap must be highly anisotropic and could possess nodes in this doping regime.
Our results demonstrate that $1/\lambda^2(T\rightarrow 0)$ varies
roughly quadradically with the superconducting transition temperature
T$_c$. We find that the superfluid density divided by the effective
mass $n_s/m^*\propto1/\lambda^s$ decreases as normal state charge
carriers are added. This implies that a form of electronic phase separation (either in real or reciprocal space) occurs
in these systems. We observe a paramagnetic frequency shift in all specimens below T$_c$, the magnitude of which is roughly independent of field but decreases with increasing doping.
\section{acknowledgments}
We appreciate the hospitality of the TRIUMF Centre for Molecular and Materials Science where the majority of these experiments were performed. Research at McMaster University is supported by NSERC and CIFAR. Work at Columbia was supported by NSF-DMR-0502706 and NSF-DMR-0806846.
Work at Ames Laboratory was supported by the Department of Energy, Basic Energy Sciences under Contract No.
DE-AC02-07CH11358. Work at the University of Maryland was supported by
AFOSR-MURI Grant No. FA9550-09-1-0603.

\end{document}